• Article •

# CDSE-UNet: Enhancing COVID-19 CT Image Segmentation with Canny Edge Detection and Dual-Path SENet Feature Fusion

Jiao Ding [1], Jie Chang [2], Renrui Han [2], Li Yang [2*]

1. School of Electrical and Electronic Engineering, Anhui Institute of Information Technology, Wuhu 241199, China.

2. School of Medical Information, Wannan Medical College, Wuhu 241000, China.

* Corresponding author，yangli@wnmc.edu.cn

**Received:**

**Supported by** the Natural Science Foundation of Anhui Province of China under Grants 2108085MF205, the Famous Teacher Training Project of Anhui Institute of Information Technology of China under Grants 22xjmspyjxmsgg02.

**Abstract Background** Accurate segmentation of COVID-19 CT images is crucial for reducing the severity and mortality rates associated with COVID-19 infections. In response to blurred boundaries and high variability characteristic of lesion areas in COVID-19 CT images, we introduce CDSE-UNet: a novel UNet-based segmentation model that integrates Canny operator edge detection and a dual-path SENet feature fusion mechanism. This model enhances the standard UNet architecture by employing the Canny operator for edge detection in sample images, paralleling this with a similar network structure for semantic feature extraction. A key innovation is the Double SENet Feature Fusion Block, applied across corresponding network layers to effectively combine features from both image paths. Moreover, we have developed a Multiscale Convolution approach, replacing the standard Convolution in UNet, to adapt to the varied lesion sizes and shapes. This addition not only aids in accurately classifying lesion edge pixels but also significantly improves channel differentiation and expands the capacity of the model. Our evaluations on public datasets demonstrate CDSE-UNet's superior performance over other leading models, particularly in segmenting large and small lesion areas, accurately delineating lesion edges, and effectively suppressing noise.

1. **Introduction**

Corona Virus Disease 2019 (COVID-19), a highly contagious and severe respiratory disease, has emerged as a grave threat to global health. Since its outbreak in December 2019, COVID-19 has spread rapidly, leading to an exponential increase in infections. The World Health Organization (WHO) reports that, as of November 11, 2022, the global case count has exceeded 630 million, with the death toll surpassing 6.58 million. In this context, rapid and precise diagnostic methods are vital in curbing the high rates of severe infections and mortality associated with COVID-19. Common diagnostic tools include Reverse Transcription-Polymerase Chain Reaction (RT-PCR) and reverse transcription polymerase chain. However, the propensity of RT-PCR tests to yield false negatives necessitates supplementary diagnostic measures, such as chest imaging[1]. The "Novel Coronavirus Infection Diagnosis and Treatment Plan (Trial Tenth Edition)" highlights key diagnostic indicators, including bilateral multifocal ground-glass opacities and pulmonary consolidation, observable via chest imaging, predominantly Computed Tomography (CT)[2]. Nonetheless, the reliability of diagnostic outcomes is often contingent on the interpreting physician's expertise, leading to potential variability in diagnoses[3,4].

Accurate segmentation of lesion areas in COVID-19 CT images is imperative in mitigating the severity and mortality of the disease. However, the pandemic's rapid spread and the sheer volume of patients render traditional manual diagnosis methods inadequate. The advent of computer technology, particularly the application of machine learning algorithms like SVM[5] and decision trees[6], initially marked advancements in medical image segmentation. Despite these developments, their efficacy remained limited. The recent surge in deep learning technologies, especially Convolutional Neural Networks (CNNs)[7], has significantly outperformed traditional algorithms in feature extraction and robustness. Fully Convolutional Networks (FCNs)[8], as seminal works in image semantic segmentation, utilize upsampling in their final layer to achieve pixel-level classification. However, the simplistic upsampling approach in FCNs often results in the loss of crucial shallow semantic features, affecting segmentation accuracy, albeit still surpassing traditional algorithms.

In leveraging the full potential of CNN layers' semantic features, Ronneberger et al. conceptualized the UNet model[9]. This architecture, shaped like the letter 'U', comprises 4 encoding layers, 2 bottleneck layers, and 4 decoding layers, with interconnected encoding and decoding layers. UNet's structured network, requiring minimal sample training, has substantially enhanced medical image segmentation performance.

Given UNet's impactful role in medical image segmentation, numerous refinements have been proposed, falling into two broad categories: structural enhancements and performance optimization. Structural enhancements generally focus on modifying encoders and decoders, as seen in variants like Res-UNet[10], Dense-UNet[11], Attention-UNet[12], Trans-UNet[13], and MltiResUNet[14]. Wang et al.[15] introduced CopleNet, a UNet-like model for segmenting COVID-19 CT image lesions, employing bridging layers and a noise-robust DSC loss function to enhance noise resilience.

UNet and its evolved variants can autonomously segment lesions in COVID-19 CT images, markedly improving upon conventional methodologies. Nevertheless, challenges persist in achieving optimal segmentation accuracy and robustness. These challenges stem from the significant variability in lesion sizes and shapes, coupled with the indistinct boundaries between lesions and normal tissues in COVID-19 CT images, as illustrated in Figure 1. To address these issues, some researchers have amalgamated image edge detection features into CNN models to bolster edge feature extraction, thereby facilitating object edge pixel classification. Bertasius G et al.[16] introduced the DeepEdge model, employing the Canny operator for initial edge information extraction, followed by semantic fusion with corresponding network features of varying sizes to enhance edge pixel classification accuracy. To tackle the diverse lesion area sizes, inspiration was drawn from the Inception network[17], leading to the design of convolutional kernels of multiple sizes to extract semantic features at different granularities. For instance, Dolz et al.[18] devised the Dense Multi-path U-Net model to accommodate the high variability in ischemic stroke lesion characteristics, using 3×3, 5×5, and 7×7 convolutional kernels for multi-scale semantic information extraction, thereby boosting model robustness.

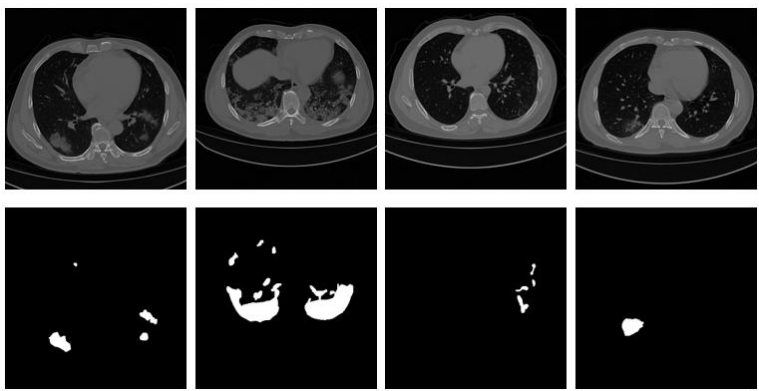

**Figure 1 Illustrative examples of COVID-19 CT images and their corresponding segmentation masks. The first row shows the original CT images of patients infected with COVID-19, capturing the varied manifestations of lung lesions. The second row presents the corresponding segmentation masks, delineating the diverse nature of these lung lessions.**

In addressing the challenges associated with the substantial variability in lesion sizes and shapes in COVID-19 CT images, as well as the blurred boundaries with normal regions, this paper proposes the CDSE-UNet model. This model, predicated on the Canny operator and dual-path SENet feature fusion, offers the following contributions:

- The use of the Canny operator to generate edge detection images for samples, extracting their semantic features with the identical network structure as the samples, and fusing them at corresponding network layers. This approach emphasizes edge information during training.

- The design of the Double SENet Feature Fusion Block (DSENetFFB), to improve channel differentiation, suppress irrelevant features, and emphasize relevant ones. This module is utilized in two scenarios: merging edge detection features with the image's inherent features, and replacing skip connections between encoders and decoders.

- The development of the Multi-Scale Convolution Block (MSCovB) within the encoder module, incorporating convolutional kernels of various sizes to maintain a balance between local and global semantic feature extraction.

## 2. Related Work

### 2.1 Image Segmentation Based on Deep Learning

Deep learning refers to artificial neural networks with deep structures and is currently the most popular machine learning technique. In 2012, Alex and others introduced a deep learning algorithm called Alex Net[7], a type of Convolutional Neural Network (CNN). In that year's ImageNet competition, it significantly outperformed traditional machine learning algorithms like SVM[5] and decision trees[6], marking the beginning of the deep learning boom. In 2015, Jonathan Long and colleagues were the first to apply deep learning to the field of image segmentation. Their design, known as Fully Convolutional Networks (FCN), achieved pixel-level image segmentation[8]. Since then, a plethora of image segmentation models based on deep learning have emerged.

Ronneberger and his team designed the renowned U-Net model[9]. It primarily consists of 4 encoder layers, 2 bottleneck layers, and 4 decoder layers. The network structure resembles the uppercase letter 'U'. The encoder is connected to the corresponding decoder layer in the network structure. The encoder comprises convolution layers, BN (Batch Normalization), RELU (Rectified Linear Unit), and downsampling. In contrast, the decoder consists of convolution layers, BN, RELU, and upsampling. The bottleneck layers contain convolution layers and RELU. Thanks to the model's encoder-decoder structure and skip connections, it can achieve semantic segmentation of images with a limited amount of training data. The model has been successfully applied to the field of medical image segmentation.

Given the notable performance of the U-Net model in medical image segmentation, numerous improved U-Net algorithms have emerged, such as Res-UNet[10], DenseUNet[11], Attention-UNet[12], Trans-UNet[13], and MultiResUNet[14], among others. Res-UNet was designed for retinal vessel segmentation tasks. It introduces skip connections into the UNet architecture and applies attention to parts of the eyeball images, increasing the network's depth and focus on crucial image regions. Dense-UNet, inspired by dense connections, replaces the UNet's encoder module with a densely connected module. In the dense connection module, subsequent layers are formed by concatenating all preceding layers. Multi-level connections facilitate the transmission of subtle image features. This model was initially used for the removal of artifacts in sparse photoacoustic tomography. Attention-UNet is a type of UNet that incorporates a gate-controlled attention mechanism. The network replaces skip connections between the encoder and decoder with Attention Gates, emphasizing essential features and suppressing irrelevant ones. MltiResUNet modifies the size of the convolution module, replacing the original 3x3 convolution kernel with a combination of 3×3, 5×5, and 7×7, to extract image features from multiple resolutions. Additionally, the network replaces the skip connections in U-Net with residual connections.

## 2.2 Edge Detection

Similar to image features like shape, color, and texture, edges are among the most fundamental characteristics of an image and play a vital role in effective image analysis. An edge refers to the discontinuity in local properties of an image, such as sudden changes in grayscale or structure. Edge detection is a common digital image segmentation technique. By analyzing the differences in pixel grayscale values within an image, one can pinpoint areas of brightness variation at object boundaries, thereby extracting edge or contour information of objects within the image.

Currently, common edge detection algorithms include the Sobel operator[19], Prewitt operator[20], Roberts operator[21] and Canny operator[22]. The Roberts operator utilizes a local differential operator to detect edges in an image. While it offers good edge localization, it may result in the loss of some edges. Additionally, this operator doesn't include image smoothing, leading to poor noise resistance. As such, it's best suited for images with sharp edges and minimal noise. Both the Sobel operator and the Prewitt operator apply weighted smoothing to neighboring pixel points in an image, followed by a differentiation operation. However, the two operators use different weights for smoothing. They both provide decent noise resistance and good edge localization, but they might detect false contours in the edges[1]. The Canny operator uses high and low thresholds to detect strong and weak edge points in an image, respectively. Based on pre-established pixel connection rules, weak edge detection points are linked to strong edge detection points, ensuring the most accurate representation of actual object edges in the image with good edge localization. These detection operators identify the discontinuity in grayscale or gradient values of image pixels and then adjust the grayscale threshold or gradient values to detect the most prominent edges or contours of objects.

## 2.3 Channel Attention Mechanism

The attention mechanism is a technique to reinforce local information and has been extensively incorporated into various deep learning models. By adding trainable attention weights to the network model, the model can automatically adjust weight parameters during training. This enhances the weights of key regions while suppressing those of irrelevant areas, allowing the model to mimic human cognitive functions and better accomplish its tasks[23].

Attention mechanisms can be categorized into channel attention, spatial attention, and hybrid attention based on their processing objects. Spatial attention focuses on feature map-level features, channel attention centers on channel-level features, while hybrid attention simultaneously concentrates on both feature map and channel-level features.

Hu introduced a channel attention mechanism called SENet (Squeeze-andExcitation Networks)[24], which won the ImageNet competition that year. SENet considers the relationships between feature channels and incorporates attention mechanisms on them. SENet automatically learns the importance of each feature channel and uses the derived importance to enhance relevant features while suppressing less important ones for the current task. This is achieved through the Squeeze and Excitation modules. Algorithms improved upon SENet include ECANet (Efficient Channel Attention Network)[25], GCNet (Global Context Network)[26], and DANet (Dual Attention Network)[27]. ECANet removes the Squeeze and Excitation operations from SENet, substituting them with local cross-channel interactions, which further reduces the model's parameter count compared to SENet. GCNet introduces the Global Context block to obtain global information relationship vectors, providing a broader global perspective compared to SENet. DANet designs a network that merges both spatial and channel attention, with channel attention being implemented through the Position Attention Module.

## 3. Materials and Methods

The CDSE-UNet model, utilizing UNet as its foundational architecture, is depicted in Figure 2. In this innovative approach, each encoder layer receives two distinct inputs: a Canny operator-based edge detection image and the original sample image. These inputs, sharing a uniform network structure, are processed using the Multi-Scale Convolution Block (MSCovB), which replaces the conventional Convolution in Unet, for enhanced feature extraction. The resulting features, with their channel numbers doubled yet maintaining the same window size, are fused using the Double SENet Feature Fusion Block (DSENetFFB) based on SENet. This fusion maintains the channel number and window size consistent. However, the origins of the two inputs differ in subsequent layers. For the sample image input, features generated from the previous layer's merged inputs through DSENetFFB are utilized, while the Canny edge detection image input is derived from its own convolved features. Transition between layers is achieved via Pooling layers, preserving the feature channel numbers while halving the window sizes.

The model's bottleneck layer and decoder structure mirror that of the standard UNet network. The key distinction lies in the feature extraction process, conducted via MSCovB, and in the replacement of UNet's skip connections with DSENetFFB. The subsequent sections delve into the intricate details of the CDSE-UNet implementation.

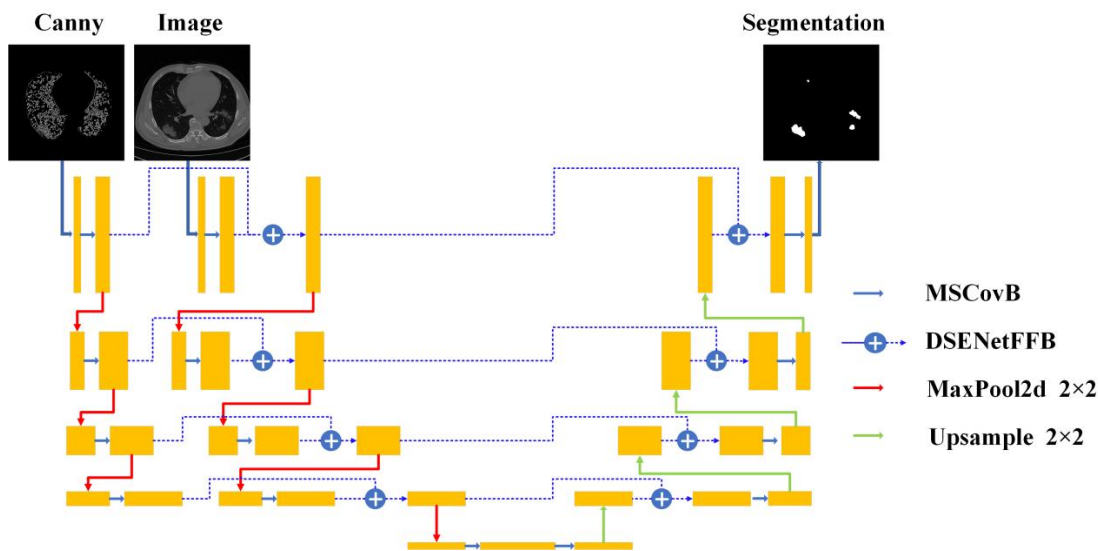

**Figure 2** The architecture of the proposed CDSE-UNet integrates Canny edge detection and Double SENet Feature Fusion Blocks (DSENetFFB) into a UNet framework. DSENetFFB as a key component enhances the feature fusion process across different network layers. MSCovB stands for Multi-Scale Convolution Block, which facilitates the extraction of detailed semantic features across various scales and sizes, crucial for handling the diverse lesion characteristics in COVID-19 CT images.

### 3.1 Edge Detection Image Based on Canny Operator

Lesion areas in COVID-19 CT images, often manifesting as ground-glass opacities and infiltrative shadows, are marked by blurred boundaries with normal regions. The precision of edge features in these areas is critical for segmentation accuracy. Drawing inspiration from[16], this study integrates edge detection images into the COVID-19 CT image segmentation model to enhance the classification of edge pixels in lesion areas. Edges, akin to shape, color, and texture features, are fundamental to image analysis, representing discontinuities in local image properties like abrupt changes in grayscale or structure. Image edge detection, a prevalent technique in digital image segmentation, identifies object edge brightness variations by analyzing pixel grayscale differences. Prominent edge detection algorithms include Sobel, Prewitt, Roberts, and the pivotal Canny operators. The Canny operator, particularly, excels in detecting brightness changes through differential operations, distinguishing strong and weak edge points using high and low thresholds. This operator is known for its low edge detection error rate and superior edge positioning accuracy, outshining other operators in numerous applications.

### 3.1.1 Fusion of Edge Detection and Sample Images

In image segmentation, the supervisory role of edge detection images is leveraged by directly stacking the features of the original sample image with those of the edge detection image, as visualized in Figure 3. However, these images, belonging to distinct modalities, possess complex intermodal semantic relationships as elucidated by Srivastava et al.[28]. Simple multimodal stacking could overlook these intricate correlations. Inspired by[18], which proposed multimodal image fusion paths for semantic integration at various levels, this study adopts an image feature fusion method illustrated in Figure 2. This method, treating the original sample image and edge detection image as dual inputs with identical semantic feature extraction network structures, fuses them at each network layer, thereby enhancing the correlation between the two image types and facilitating accurate classification of image edge pixels.

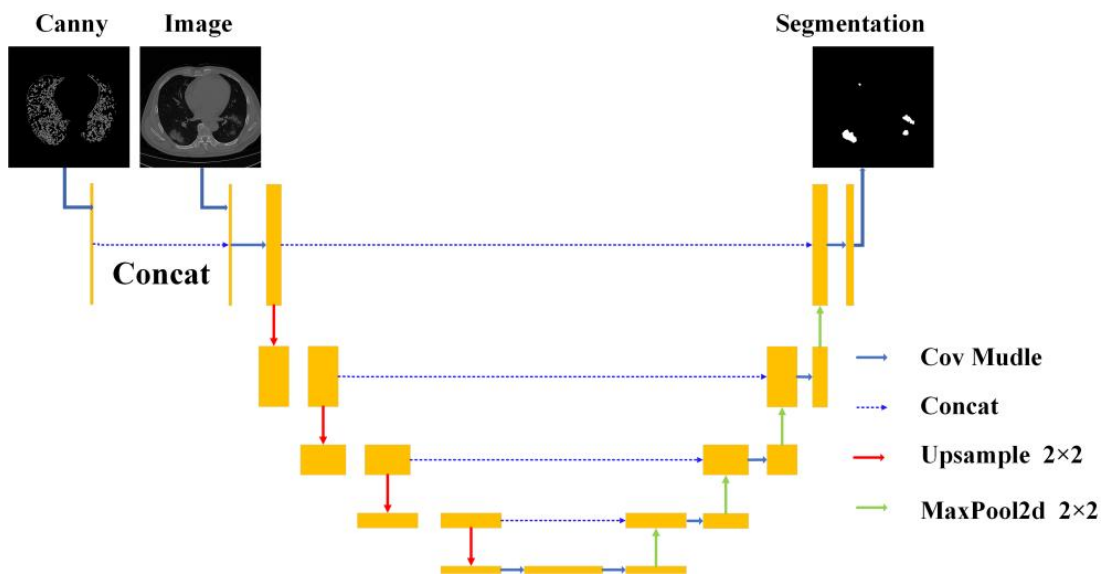

**Figure 3** The process of feeding the CDSE-UNet model with a combined input of Canny edge-detected images and original CT scans. However, these two types of images possess complex multi-modal semantic relationships, and a simple stacking approach might overlook the correlations between them, thereby diminishing the supervisory role of the edge detection image.

### 3.1.2 Generation of edge detection image based on Canny operator

Following the methodology outlined in[29], the Canny operator is employed to extract edge features. The process involves several steps:

1. Gaussian filtering. Gaussian filtering primarily smoothens the image, suppressing noise within it. Specifically, a two-dimensional Gaussian kernel of a particular size convolves with the image. The Gaussian kernel $G(x,y)$ is presented in Eq. (1).

$$G(x,y) = \frac{1}{2\pi\sigma^2} \exp(-\frac{x^2+y^2}{2\sigma^2}) \tag{1}$$

2. Calculating gradient magnitude and direction for pixel points. The Sobel first-order derivative operator is used to calculate the gradient magnitude of each pixel point in the $x$ and $y$ directions. The Sobel operator is shown in Eq. (2).

$$S_x = \begin{pmatrix} -1 & 0 & 1 \\ -2 & 0 & 2 \\ -1 & 0 & 1 \end{pmatrix}, S_y = \begin{pmatrix} 1 & 2 & 1 \\ 0 & 0 & 0 \\ -1 & -2 & -1 \end{pmatrix} \quad (2)$$

Using $S_x$ and $S_y$, calculating gradient matrices, $G_x$ and $G_y$, for grayscale image $I$ in the $x$ and $y$ directions, respectively, as shown in Eq. (3).

$$G_x = S_x * I, G_y = S_y * I \quad (3)$$

The image gradient matrix $G_{xy}$ is then calculated based on the formulation $g_{xy}(i,j) = \sqrt{g_x(i,j)^2 + g_y(i,j)^2}$

3. Non-maximum suppression of pixel gradient. An 8-neighborhood is constructed for pixel points. Based on the sign and magnitude of gradient strengths in the $x$ and $y$ directions, the gradient direction region is determined. Then, using the gradient strength values of adjacent pixel points, gradient values for comparison points in the positive and negative directions are calculated using linear interpolation, as shown in Eq. (4) and (5).

$$g_{up}(i,j) = (1-t) \cdot g_{xy}(i, j+1) + t \cdot g_{xy}(i-1, j+1) \quad (4)$$

$$g_{down}(i,j) = (1-t) \cdot g_{xy}(i, j-1) + t \cdot g_{xy}(i+1, j-1) \quad (5)$$

Where $t = |g_y(i,j)/g_x(i,j)|$. The gradient values of the comparison points are compared with the gradient strength of the pixel point. If the gradient value of the pixel point is maximal, the pixel point is retained; otherwise, it is suppressed.

4. Threshold hysteresis. A gradient strength threshold range [*low*, *high*] is set. The retained pixel point gradient strength *k* is compared with thresholds, *low* and *high*, to determine if it should be considered an edge point in the image. There are three scenarios: (1) When *k* > *high*, it is retained. (2) When $low \leq k \leq high$, its status is pending. (3) When *k* < *low*, it is eliminated.

5. Elimination of isolated weak edges. Pixel points with gradient strength within the range [*low*, *high*] are marked as pending, meaning they are potential edge points that need to be checked against their 8 neighboring pixels. If any of these neighboring pixels have been retained, the pending pixel is preserved as an edge point; otherwise, it is eliminated.

## 3.2 Dual-Path Feature Fusion

The network model designed for the fusion of edge detection and sample images ensures efficient fusion of features from both inputs at every network layer. Feature fusion is initially achieved through channel stacking, as depicted in the Simple Concatenation Block in Figure 4(a). This approach treats all channels equally, yet the significance of each channel varies. The channel attention mechanism, as exemplified by SENet (see Figure 5), adeptly represents channel weights, suppressing irrelevant and enhancing important features. The mechanism involves global average pooling, followed by Squeeze and Excitation operations, and finally, Sigmoid normalization, leading to the generation of channel attention parameters which are multiplied element-wise with the input feature map to yield the output feature map.

Building on Simple Concatenation, the Single SENet Feature Fusion Block, as shown in Figure 4(b), was introduced, employing SENet to augment feature fusion efficiency. Given the distinct modalities and channel relationships of the edge detection and sample images, the Double SENet Feature Fusion Block was conceptualized. This block, shown in Figure 4(c), first selects channel attention before merging features, thus accommodating the modalities' differences and enhancing channel differentiation, as evidenced by subsequent experimental results. In the UNet model, skip connections facilitate the addition of shallow semantic features from the encoder to the upsampled features in the decoder. Accordingly, the Double SENet Concatenation Block, a suitable replacement for these connections, was integrated into our network model.

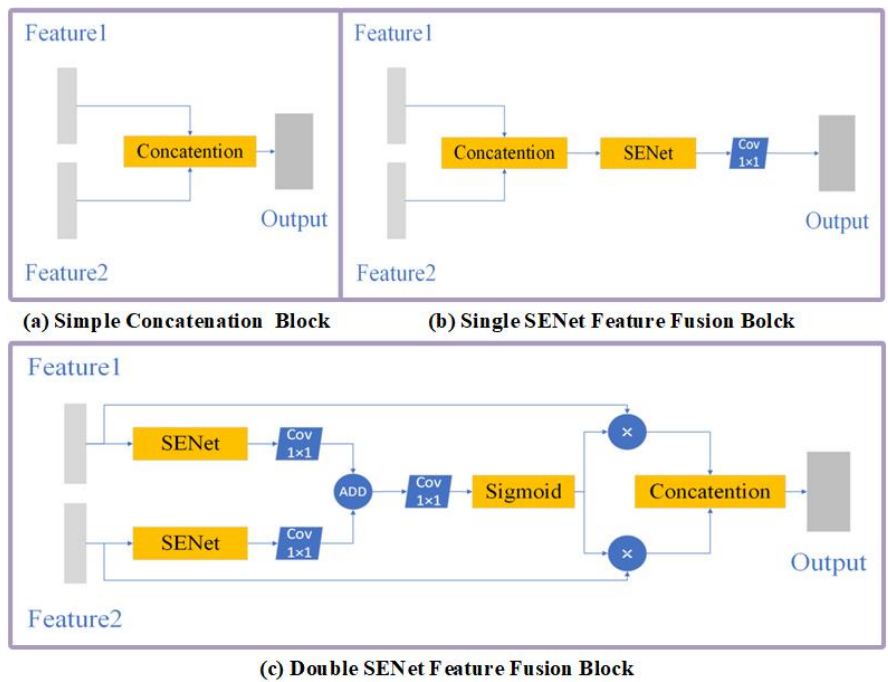

Figure 4 A visual comparison of three distinct feature fusion approaches. (a) displays the simple Concatenation block, where features are straightforwardly stacked, providing a basic level of fusion. (b) illustrates the single SENet feature fusion block, which introduces the SENet mechanism for more nuanced channel-wise feature integration. (c) depicts the double SENet feature fusion block our model

adopts, an advanced fusion approach that employs dual SENet structures for deeper and more effective feature integration from different image modalities.

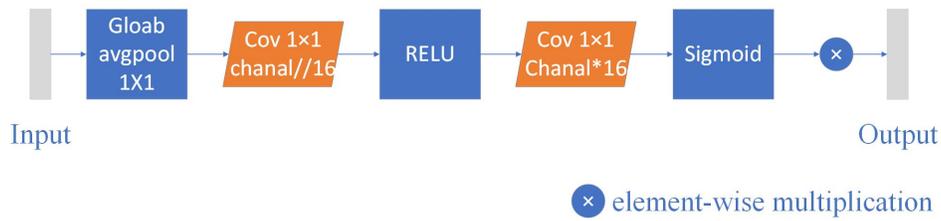

Figure 5 The Squeeze-and-Excitation Networks (SENet) architecture, which focuses on adaptively recalibrating channel-wise feature responses by explicitly modelling interdependencies between channels.

### 3.3 Multi-Scale Convolution Block

COVID-19 patients typically exhibit varying lesion sizes and shapes across the incubation, progression, and recovery phases. The model's capability to segment lesions of different sizes is essential, particularly in the early detection stages to mitigate the risk of worsening conditions. The convolution kernel size in the model determines the receptive field of the input feature image. The UNet model's fixed 3 × 3 kernel size is inadequate for the diverse lesion areas seen in COVID-19. The Inception module in the InceptionNet network, comprising variously sized convolution kernels, addresses this limitation by broadening the network model.

This paper incorporates a multi-size convolution kernel module within the network model, enhancing the extraction of both local and global lesion area features. The module includes 1 × 1, 3 × 3, 5 × 5, and 7 × 7 sized convolutions. Initially designed as depicted in Figure 6(a), the model adopted in this paper, illustrated in Figure 6(b), reduces computational demands while maintaining effective feature extraction.

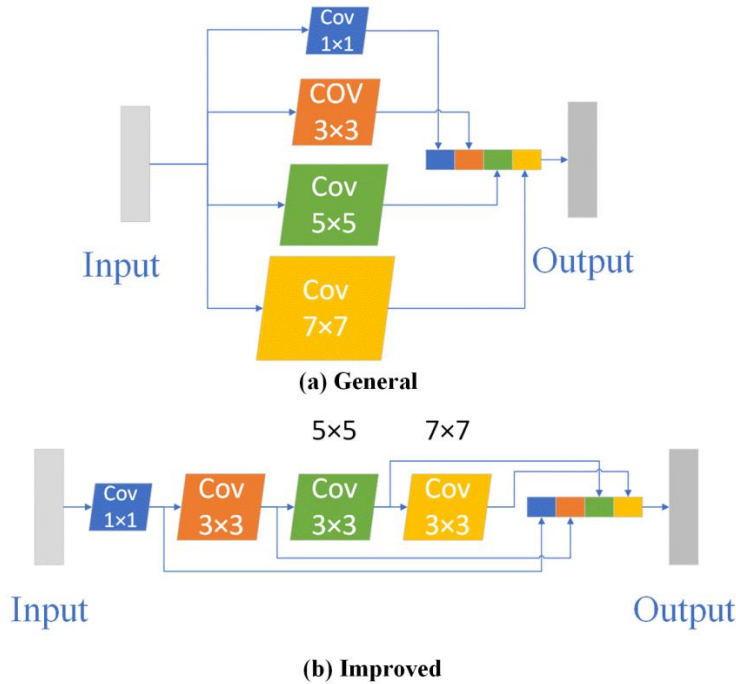

Figure 6 The multi-size convolution kernel module. To address the diverse and complex patterns found in COVID-19 CT images, the architecture of the Multi-Scale Convolution Block (MSCovB) consists of convolutional layers of multiple kernel sizes such as 1 × 1, 3 × 3, 5 × 5, and 7 × 7, in a parallel configuration. This setup allows the block to simultaneously process and combine information from both local and global perspectives.

### 3.4 Loss Function

This paper combines Binary Cross Entropy (BCE) and Dice loss as the loss function as following.

$$Lf = 0.5*BCELoss + 0.5*DiceLoss \qquad (6)$$

## 4. Experiment

### 4.1 Experimental Environment

The software development environment used was PyCharm Community Edition 2021 with the development framework being PyTorch. The primary configuration environment includes python3.7, torch1.8.1, cuda111, numpy1.20.3, and others. The hardware environment was Nvidia GeForce GTX 3070 (8 GB). The primary parameters for model training include: the Adam optimizer, an initial learning rate of 0.001, dynamic learning rate adjustment every 30 epochs with a decay factor of 0.9, a maximum of 300 training iterations, and retaining the model with the highest Dice Similarity Coefficient (DSC) value on the test set as the optimal model.

### 4.2 Datasets

The experimental data is sourced from image slices in the public dataset "COVID-19 CT lung and infection segmentation dataset"[30]. This dataset comprises 20 cases, categorized into 'coronacases' and 'pneumonia'. Experienced radiologists annotated and validated the images, with three types of annotations: left lung, right lung, and the infection area. Since most of the image slices did not display infection areas, this study selected 878 slices that predominantly featured infection areas. From these, 788 were randomly chosen as the training set and 90 for the test set. This paper focuses on segmenting the infected areas of the lungs, retaining only the infection mask data. Some data samples and their corresponding masks are shown in Figure 1.

### 4.3 Evaluation Metrics

This study employs three commonly used image segmentation metrics for four quantitative assessments: Accuracy, Precision, Recall, and Dice Similarity Coefficient (DSC). Accuracy evaluates the correctness of predicted pixel points. Precision assesses the accuracy of pixels predicted as positive samples. Recall measures the proportion of correctly predicted positive samples out of all positive samples. The Dice Similarity Coefficient evaluates the overlap between two sets, assessing the similarity between two images. All metrics range between 0 and 1, with performance being better as they approach value 1.

## 5. Discussion and Ablation Experiments

### 5.1 Discussion

This study compared the lung CT image segmentation network, CDSE-UNet, proposed in this paper with advanced medical image segmentation models such as UNet, Attention-UNet, Swin-Unet [31], Trans-UNet, and Dense-UNet. All experiments were conducted under the same hardware environment, and training parameters like learning rate, batch size, loss function, max pooling, and up-sampling were kept consistent across all models. The number of epochs was uniformly set to 300 for all models, and the highest DSC value on the test set was chosen as the experimental result for each model. Detailed metrics for each experiment are presented in Table 1.

Table 1 Comparison of evaluation metrics for different methods

| Method | Accuracy | Precision | Recall | DSC |
| --- | --- | --- | --- | --- |
| UNet[9] | 0.9925 | 0.7803 | 0.9369 | 0.8917 |
| Attention-UNet[12] | 0.9925 | **0.8216** | 0.9602 | 0.8898 |
| Trans-Unet[13] | 0.9917 | 0.8128 | 0.9011 | 0.8657 |
| Swin-Unet[31] | 0.9918 | 0.7972 | 0.9235 | 0.8725 |
| Dense-UNet[11] | 0.9924 | 0.8197 | 0.9418 | 0.8943 |
| **CDSE-UNet** | **0.9930** | 0.8135 | **0.9648** | **0.9107** |

Table 1 shows that the CDSE-UNet model achieves the best values in terms of Accuracy, Recall, and DSC metrics, especially showing significant advantages in the Recall and DSC metrics. In terms of the DSC metric, CDSE-UNet reaches 91.07%, making it the only model exceeding 90%, and shows an improvement of 1.66% compared to the second-best model, Dense-UNet. For the Recall metric, CDSE-UNet improves by 0.58% compared to the next best model, Attention-UNet. In the Accuracy metric, given that this task is a binary classification with background pixels easily classified correctly and substantially outnumbering infection area pixels, all models score high with minimal differences. Nevertheless, CDSE-UNet still performs the best, surpassing the next best model, Attention-UNet, by 0.06%. However, in the Precision metric, CDSE-UNet falls short by 0.68% compared to the best-performing model, Attention-UNet. CDSE-UNet exhibits a significant advantage in the crucial DSC metric, demonstrating the strongest ability in precise segmentation of lesions.

Figure 7 shows the segmentation comparison of CDSE-UNet with other advanced models on five CT images. The images corresponding to "Image" are the original images, those corresponding to "Label" are the segmentation labels, and the images corresponding to the model names represent the segmentation results of each model. CDSE-UNet has obvious advantages in segmenting large lesion areas, small lesion areas, lesion edges, and in suppressing noise points. In the first and second images from the left in the figure, the lesion areas vary in size, and the lesions segmented by CDSE-UNet are the closest to the Label image. In the third image from the left, there are five lesions, one large and four small, in the left lung. Only CDSE-UNet segmented all five lesions, and the lesions segmented by CDSE-UNet have shapes that are more similar to the Label image. In the fourth and fifth images from the left, the lesion edges segmented by CDSE-UNet are closer to the Label image. In summary, CDSE-UNet places more emphasis on the details of the lesion edge while suppressing noise points, achieving impressive visual segmentation results.

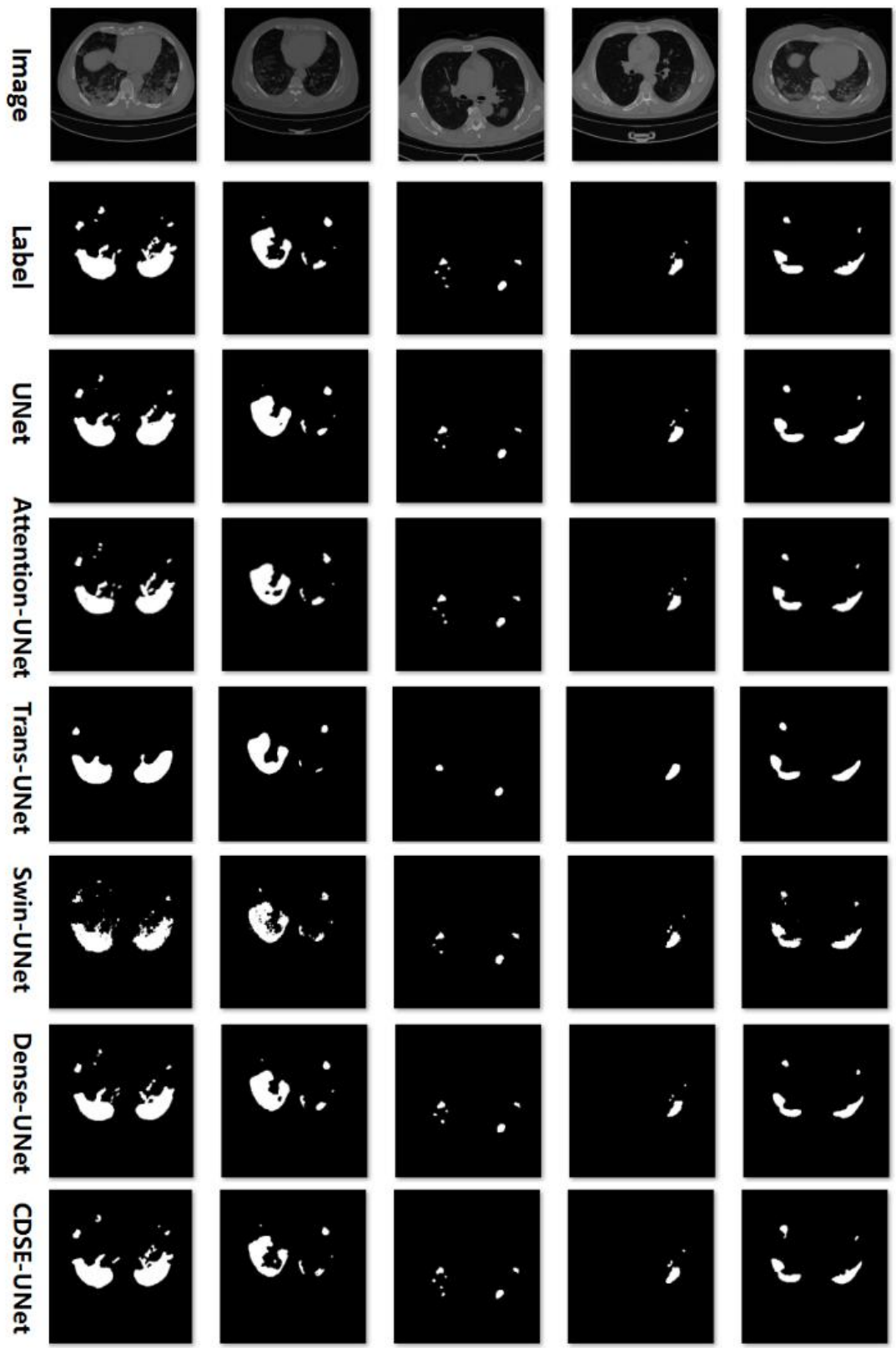

**Figure 7 The segmentation comparison of CDSE-UNet with other counterpart models on five CT images.**

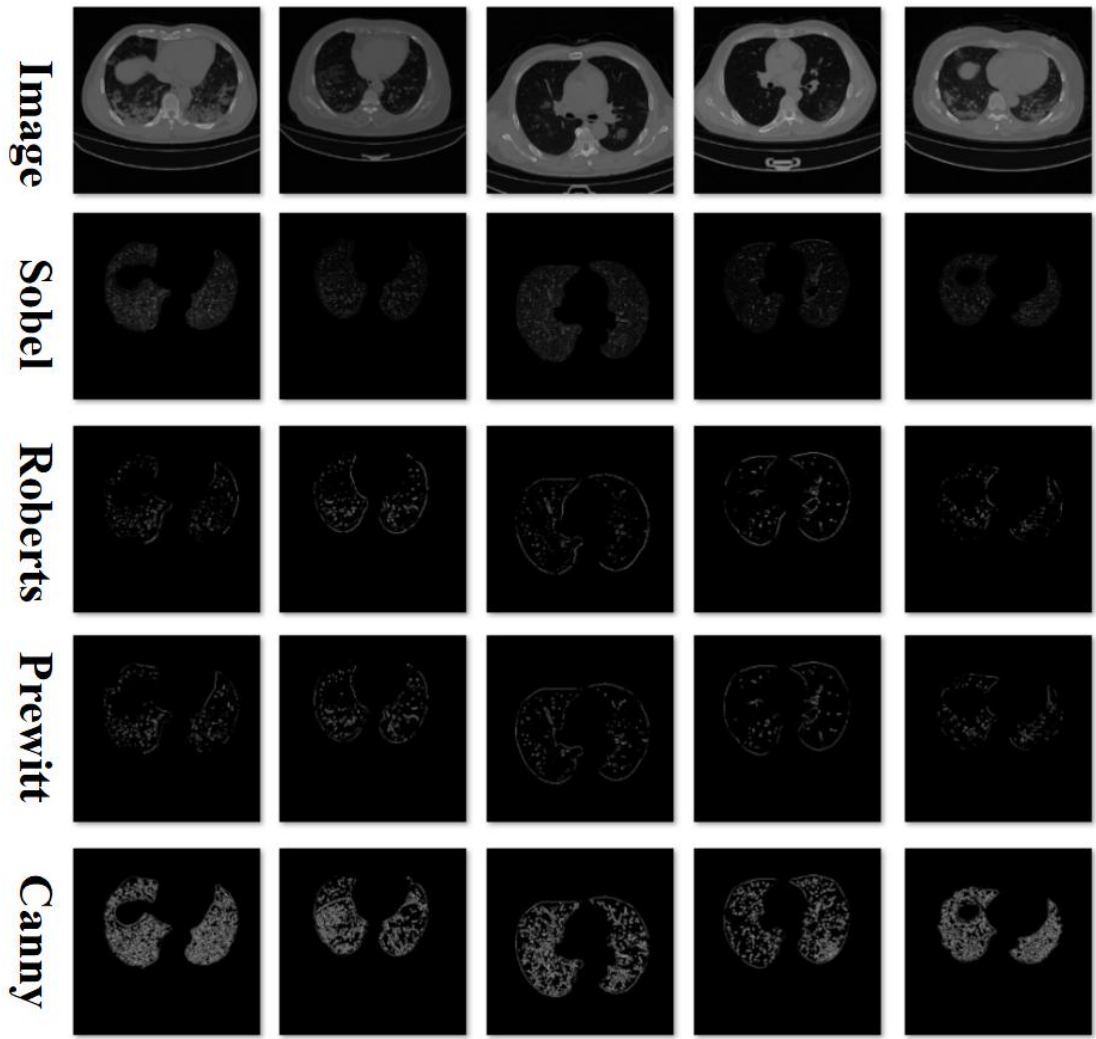

Figure 8 The comparison of four models in ablation experiment.

**5.2 Ablation Experiments**

**5.2.1 Comparative Experiment of Different Edge Detection Operators**

The edges of an image contain a wealth of feature information, and edge detection operators are used to extract this edge information. Inspired by Swin-UNet, this study uses various edge detection algorithms to process sample data, generating sample edge images. These images are merged with the samples via channel connections and fed into the model for training. This approach makes the model focus more on edge features, which can enhance the performance of image segmentation. Different edge detection operators perform differently in various scenarios. To find an operator suitable for segmenting the edges of infection areas in COVID-19 CT images, we compared the Prewitt, Roberts, Sobel, and Canny edge detection operators. Figure 6 shows the results of these four operators on five COVID-19 CT images, with a graythresh value of 0.2. From Figure 8, it can be observed that, compared to the other three operators, the Canny operator detects more image edge details, and its edges are also smoother.

Table 2 Comparison of evaluation metrics among

| Method | Accuracy | Precision | Recall | DSC |
|---|---|---|---|---|
| Sobel&CDSE-UNet | 0.9928 | **0.8271** | 0.9483 | 0.9062 |
| Roberts&CDSE-UNet | 0.9927 | 0.8016 | 0.9421 | 0.8999 |
| Prewitt&CDSE-UNet | 0.9928 | 0.8237 | 0.9621 | 0.9071 |
| Canny&CDSE-UNet | **0.9930** | 0.8146 | **0.9648** | **0.9107** |

Table 2 compares the evaluation metrics of the CDSE-UNet network combined with each of the four edge detection operators. The combination of the Canny operator with CDSE-UNet outperforms the combinations of other operators with CDSE-UNet in terms of Accuracy, Recall, and DSC metrics.

### 5.2.2 Comparative Experiment on Different Feature Fusion Methods

In Section 3.2, we described three different methods of channel feature fusion, which are used to merge the semantic features of image edge detection, the image's own semantic features, and the skip connections between the encoder and decoder. These three methods are simple channel stacking, single attention mechanism channel stacking, and double attention mechanism channel stacking, corresponding to the network structures shown in Figure 4(a), Figure 4(b), and Figure 4(c), respectively. To validate the effectiveness of different feature fusion methods, an ablation study was conducted, and the results are shown in Table 3. As can be seen from the table, the third method, Double SENet Feature Fusion Block, achieved the highest values in Accuracy, Precision, Recall, and DSC, confirming the effectiveness of this feature fusion approach.

Table 3 Comparison of three different channel feature fusion methods

| Feature Fusion Method | Accuracy | Precision | Recall | DSC |
|---|---|---|---|---|
| Simple Concatation | 0.9927 | 0.8037 | 0.9617 | 0.9035 |
| Single SENet Concatation | 0.9928 | 0.8085 | 0.9629 | 0.9077 |
| Double SENet Concatation | **0.9930** | **0.8146** | **0.9648** | **0.9107** |

### 6. Conclusion

The segmentation of COVID-19 CT infection areas serves as a quantitative diagnostic basis for physicians, aiding in reducing diagnostic time and enhancing accuracy. The proposed CDSE-UNet model primarily incorporates the fusion of Canny operatorbased sample edge detection features and the corresponding fusion method, the Double SENet Feature Fusion Block. The former enables the model to focus more on lesion edge pixels, while the latter enhances the differentiation of channel features and the representational capacity of important features. Additionally, the model integrates a multi-scale convolution module, effectively balancing the extraction of both local and global features. Extensive comparative experiments demonstrate that CDSE-UNet pays greater attention to lesion edge details, yielding impressive segmentation results. However, numerous edge detection methods exist, and this study has so far only compared four such operators—Prewitt, Roberts, Sobel, and Canny. Future research will explore more edge detection operators and their improvements to further enhance segmentation performance.

**Declaration of competing interest**

We declare that we have no conflict of interest.

**Abbreviations**

The following abbreviations are used in this manuscript:

- **BCE**:          Binary Cross Entropy
- **BN**:           Batch Normalization
- **CDSE-UNet**: Enhancing COVID-19 CT Image Segmentation with Canny Edge Detection and Dual-Path SENet Feature Fusion
- **CNN**:          Convolutional Neural Network
- **DSC**:          Dice Similarity Coefficient
- **DSENetFFBt**: Double SENet Feature Fusion Block
- **FCN**:           Fully Convolutional Networks
- **MSCovB**:    Multi-Scale Convolution Block
- **ReLU**:         Rectified Linear Unit
- **SENet**:        Squeeze-and-Extend Networks
- **SVM**:          Support Vector Machine
- **UNet**:          Neural Networks with U-Net Architecture for Image Segmentation